%% file: SugraInfBK18_5.tex
\documentclass[%
superscriptaddress,
nofootinbib,
amsmath,amssymb,
prl,
twocolumn,
]{revtex4-2}

\usepackage{graphicx}% Include figure files
\usepackage{dcolumn}% Align table columns on decimal point
\usepackage{bm}% bold math
\usepackage{amsfonts}
\newcommand{\C}[1]{{\mathcal #1}}

\newcommand{\BS}[1]{{\boldsymbol #1}}

\newcommand{\half}{\frac 12}

\newcommand{\Slash}[1]{{\ooalign{\hfil#1\hfil\crcr\raise.167ex\hbox{/}}}}

\begin{document}

%%%%%%%%%%%%%%%%%%%%%%%%%%%%%%%%%%%%%%%%%%%%%%%%

%\preprint{APS/123-QED}

%%%%%%%%%%%%%%%%%%%%%%%%%%%%%%%%%%%%%%%%%%%%%%%%

\title{Gravitino constraints on supergravity inflation
}% Force line breaks with \\

\author{Shinsuke Kawai}
%\email{kawai@skku.edu}
\affiliation{Department of Physics, 
Sungkyunkwan University,
Suwon 16419, Republic of Korea}
\author{Nobuchika Okada}
%\email{okadan@ua.edu}
\affiliation{
Department of Physics and Astronomy, 
University of Alabama, 
Tuscaloosa, AL35487, USA} 
\date{\today}% It is always \today, today,


\date{\today}% It is always \today, today,

%%%%%%%%%%%%%%%%%%%%%%%%%%%%%%%%%%%%%%%%%%%%%%%%
%%%%% Abstract %%%%%

\begin{abstract}
Supergravity embedding of the Standard Model of particle physics provides 
phenomenologically well-motivated and observationally viable inflationary scenarios.
We investigate a class of inflationary models based on the superconformal framework of supergravity and discuss constraints from the reheating temperature, with the particular focus on the gravitino problem inherent in these scenarios. 
We point out that a large part of the parameter space within the latest BICEP/Keck 95\% confidence contour may have been excluded by the gravitino constraints, depending on the mass scale of the inflaton.
Precision measurements of the scalar spectral index by a future mission may rule out some of these scenarios conclusively.
\end{abstract}

%%%%%%%%%%%%%%%%%%%%%%%%%%%%%%%%%%%%%%%%%%%%%%%%

%Use showkeys class option if keyword
\maketitle

%\tableofcontents
%%%%%%%%%%%%%%%%%%%%%%%%%%%%%%%%%%%%%%%%%%%%%%%%
%%%%% Body of the Paper %%%%%

%%%%%%%%%%%%%%%%%%%%%%%%%%%%%%%%%%%%%%%%%%%%%%%%
%\section{\label{sec:Intro}
{\em Introduction.}---
%
% Supergravity inflation
Understanding the origin of cosmic inflation is an important goal of particle cosmology, and for that purpose, model building in a theory beyond the Standard Model is a promising direction of research.
In particular, supergravity embedding of the Standard Model offers a well-motivated framework;
supersymmetry allows natural gauge unification, softens the hierarchy problem and provides a natural candidate for the dark matter.
% eta problem
Realizing a realistic inflationary scenario within supergravity was once considered challenging.
The statement of this difficulty, known as the $\eta$ problem, is based on assumptions including the canonical form of the K\"{a}hler potential.
The avenues to circumvent the $\eta$ problem are now well known.
In this letter, we will be concerned with a class of supergravity inflationary models obtained by relaxing the assumption of the canonical K\"{a}hler potential.
These are the direct supersymmetric analogue of the nonminimally coupled Higgs inflation type model \cite{CervantesCota:1995tz,Bezrukov:2007ep}, which has been a focus of much attention due to its solid phenomenological origin and the excellent fit of the cosmological parameters to the measurements by the WMAP and Planck satellites.

%%%%%%%%%%%%%%%%%%%%%%%%%%%%%%%%%%%%%%%%%%
\begin{figure}
\includegraphics[width=85mm]{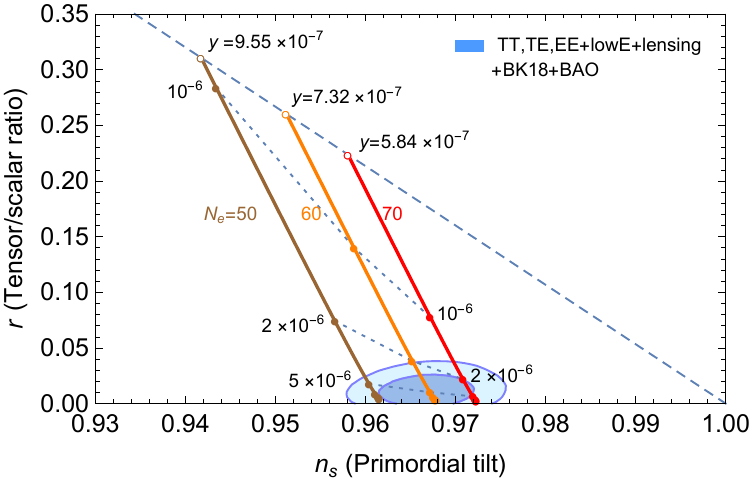}
% Here is how to import EPS art
\caption{
\label{fig:nsrYukawa}
The prediction of the primordial tilt $n_s$ and the tensor-to-scalar ratio $r$ by the nonminimally coupled supergravity inflation model, shown for the $e$-folding number $N_e=50$, $60$ and $70$.
The points for 
$y=10^{-6}$, $2\times 10^{-6}$, $5\times 10^{-6}$, $10^{-5}$, $2\times 10^{-5}$, $5\times 10^{-5}$, $1$ are marked with $\bullet$.
The thick dashed line $r=\frac{16}{3}(1-n_s)$ corresponds to the minimally coupled ($\xi=0$) model.
The blue contours on the background are the
Planck+BICEP/Keck 2018
1- and 2-$\sigma$ constraints \cite{BICEP:2021xfz}.
}
\end{figure}
%%%%%%%%%%%%%%%%%%%%%%%%%%%%%%%%%%%%%%%%%%

%%%%%%%%%%%%%%%%%%%%%%%%%%%%%%%%%%%%%%%%%%%%%%%%
%\section{\label{sec:Models}
{\em Basic structure of the supergravity inflation model.}---
The inflationary model of our interest is constructed from the supergravity lagrangian 
\begin{align}
  {\C L}\supset \int d^4\theta\, \phi^\dagger\phi\,{\C K}
  +\Big\{\int d^2\theta\phi^3 W+{\rm h.c.}\Big\},
\end{align}
%where $\phi$ is the compensating multiplet, and 
in which the superpotential is assumed to include the coupling of a singlet or adjoint superfield $S$ and a vector-like pair $(\Phi, \overline\Phi)$ under a certain gauge symmetry
\begin{align}\label{eqn:Wmess}
  W\supset y S\overline\Phi \Phi.
\end{align}
This structure is common.
Examples include the singlet $S$ and the Higgs doublet superfields $(\Phi, \overline\Phi)=(H_u,H_d)$ of the NMSSM \cite{Einhorn:2009bh,Ferrara:2010in,Ferrara:2010yw}, and $S={\BS 2\BS 4}_H$ and $(\Phi, \overline\Phi)=({\BS 5}_H,\overline{\BS 5}_H)$ of the minimal $SU(5)$ grand unification model \cite{Arai:2011nq,Kawai:2015ryj}.
See also \cite{Pallis:2011gr} for the construction in the Pati-Salam model, 
\cite{Arai:2011aa,Arai:2012em,Kawai:2014doa,Kawai:2014gqa,Kawai:2021gap} for the type I and type III seesaw models, 
\cite{Arai:2013vaa} for the $B-L$ model, 
\cite{Leontaris:2016jty} for the $SO(10)$ grand unified theory,
\cite{Okada:2017rbf} for the hybrid inflation model and
\cite{Kawai:2021tzc} for the gauge mediated supersymmetry breaking model.
The K\"ahler potential in the superconformal framework \cite{Kaku:1978nz,Siegel:1978mj,Cremmer:1982en,Ferrara:1983dh,Kugo:1982mr,Kugo:1982cu,Kugo:1983mv} is chosen in the form
\begin{align}\label{eqn:K}
  {\C K}=&-3M_{\rm P}^2+|\overline\Phi|^2 + |\Phi|^2 + |S|^2\crcr
  &\qquad -\frac 32 \gamma \left(\overline\Phi\Phi+{\rm h.c.}\right)
  -\frac{\zeta}{M_{\rm P}^2}|S|^4,
\end{align}
where $M_{\rm P}=2.44\times 10^{18}$ GeV is the reduced Planck mass, and $\gamma$, $\zeta$ are real parameters.
One may always adjust the parameter $\zeta$ so that $S$ is stabilized at some constant value, which is assumed to be small compared to the scale of inflation.
Parametrizing the scalar component of the vector-like fields along the D-flat direction as 
%\begin{align}
$\overline\Phi=\Phi=\half \varphi$,
%\end{align}
the standard supergravity computation gives the scalar part of the action
\begin{align}\label{eqn:SJ}
S_{\rm scalar}=\int d^4 x\sqrt{-g}\left[
\frac{M_{\rm P}^2+\xi\varphi^2}{2} R
-\half(\partial\varphi)^2
-\frac{y^2}{16}\varphi^4
%V_{\rm J}(\varphi)
\right].
\end{align}
Here, $\xi\equiv\frac{\gamma}{4}-\frac 16$ parametrizes the nonminimal coupling between the scalar field $\varphi$ and the scalar curvature $R$.
The action \eqref{eqn:SJ} is recognized as that of the nonminimally coupled $\lambda\varphi^4$ model \cite{Okada:2010jf} and the prediction for the cosmological parameters is obtained in the standard slow roll paradigm, after transforming it into the Einstein frame.
The {\em inflaton} field $\widehat\varphi$ canonically normalized in the Einstein frame is related to $\varphi$ by the relation
\begin{align}
  d\widehat\varphi
  =\frac{M_{\rm P}\sqrt{M_{\rm P}^2+\xi\varphi^2(1+6\xi)}}{M_{\rm P}^2+\xi\varphi^2}d\varphi.
\end{align}
The scalar potential in the Einstein frame is deformed by the factor arising from the Weyl transformation as
\begin{align}\label{eqn:VE}
  V_{\rm E}(\varphi)=\frac{y^2}{16}\frac{M_{\rm P}^4\varphi^4}{(M_{\rm P}^2+\xi\varphi^2)^2}.
\end{align}
This potential is concave for not too small $\xi$, giving the observationally supported perturbation spectrum with the suppressed tensor mode at the CMB scale.
The model has two tunable parameters $\xi$ (or $\gamma$) and $y$, but with the normalization of the scalar perturbation amplitude, there remains only one parameter degree of freedom.
%normalized as $A_s=2.xx\times 10^{-9}$ at the pivot scale $k=0.002\,\text{Mpc}^{-1}$.
As $\xi$ is increased from zero, the coupling $y$ is also increased towards a larger value.
The predicted primordial tilt $n_s$ and tensor-to-scalar ratio $r$ are shown in Fig.~\ref{fig:nsrYukawa} for different values of $e$-folding number $N_e$. 
It can be seen that $y\gtrsim 10^{-6}\sim 10^{-5}$ is in good agreement with the recent cosmological data.
Note that $y\sim 10^{-6}$ is not unnaturally small from the phenomenological perspective, as it is in the same order as the Standard Model electron Yukawa coupling.
The fact that the `self-coupling' in the potential \eqref{eqn:VE} appears as $y^2$, and not as $y$, is a salient feature of this supergravity inflation model which is in stark contrast to the nonsupersymmetric counterpart. 
For example, the Higgs inflation model \cite{CervantesCota:1995tz,Bezrukov:2007ep} requires a large nonminimal coupling $\xi\sim 10^4$ in order to accommodate the Standard Model Higgs self coupling, which led some authors to worry about the unitarity issue \cite{Barbon:2009ya,Burgess:2009ea,Hertzberg:2010dc,Lerner:2009na} (see however \cite{Bezrukov:2010jz}). 
Since the self coupling is $y^2$ in supergravity, this awkwardness, if it exists, may be easily avoided.

%%%%%%%%%%%%%%%%%%%%%%%%%%%%%%%%%%%%%%%%%%
\begin{figure*}
\includegraphics[width=120mm]{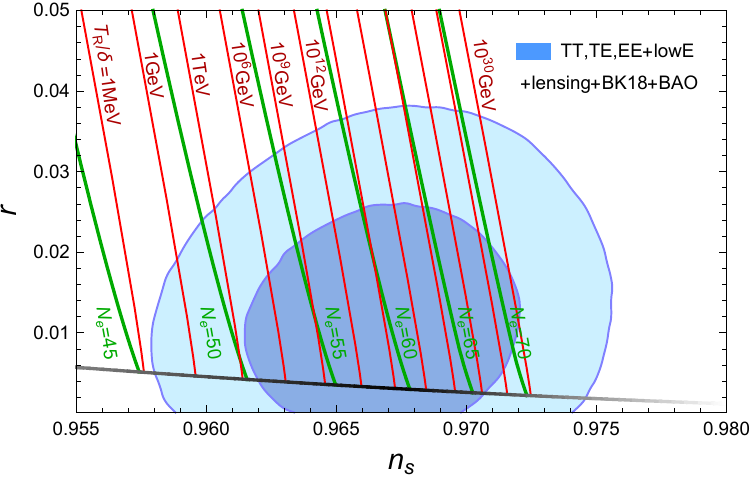}
% Here is how to import EPS art
\caption{
\label{fig:nsrTR}
The prediction for the primordial tilt $n_s$ and the tensor-to-scalar ratio $r$, computed for the rescaled reheating temperature $\delta^{-1} T_{\rm R}=1\;{\rm MeV}$, 1 GeV, 1 TeV and $10^6$, $10^9$, $10^{12}$, $10^{15}$, $10^{18}$, $10^{21}$, $10^{24}$, $10^{27}$, $10^{30}\;{\rm GeV}$ (red lines).
The curves for $e$-foldings $N_e = 45$, 50, 55, 60, 65, 70 are also indicated in green.
The near-horizontal grey curve $n_s = 1-r/4-\sqrt{r/3}$ is the prediction in the limit $\xi\gg 1$.
The contours on the background are the Planck+BICEP/Keck 2018
1- and 2-$\sigma$ constraints \cite{BICEP:2021xfz}.
}
\end{figure*}
%%%%%%%%%%%%%%%%%%%%%%%%%%%%%%%%%%%%%%%%%%

%%%%%%%%%%%%%%%%%%%%%%%%%%%%%%%%%%%%%%%%%%%%%%%%
%\section{\label{sec:GravitinoProb}
{\em Gravitino problem.}---
% Gravitino problem
Supergravity entails the gravitino, which is potentially harmful in cosmological scenarios \cite{Pagels:1981ke,Weinberg:1982zq,Khlopov:1984pf,Ellis:1984eq} depending on its mass $m_{3/2}=F/\sqrt{3}M_{\rm P}$, see e.g. \cite{Hook:2018sai}.
A stable gravitino may be produced by the decay of the inflaton, by the decay of a heavier supersymmetric particle, or thermally produced via the freeze-in mechanism. 
See \cite{Eberl:2020fml} for the details of computations of the thermal production rate.
The stable gravitino in the mass range $4.7\,{\rm eV} \lesssim m_{3/2} \lesssim 0.24\,{\rm keV}$ becomes a hot or warm dark matter component, which is severely constrained by the analysis of small scale structure formation \cite{Osato:2016ixc,Xu:2021rwg}.
In the range $0.24\,{\rm keV} \lesssim m_{3/2}\lesssim 1\,{\rm GeV}$, the gravitino behaves as cold dark matter.
The condition that the Universe is not overclosed by the gravitino sets an upper bound on the reheating temperature $T_{R}\lesssim 10^{2}\sim 10^{7}$ GeV, depending on the mass $m_{3/2}$ \cite{Moroi:1993mb}.
The gravitino in the range $1\,{\rm GeV}\lesssim m_{3/2}\lesssim 1\,{\rm TeV}$ is restricted due to light element photodestruction.
The overclosure bound for the $m_{3/2}\gtrsim 1\,{\rm TeV}$ gravitino dark matter gives $T_{\rm R}\lesssim 10^9$ GeV.
The gravitino with $m_{3/2}\geq 1\,{\rm TeV}$ is likely to be unstable.
The condition that the successful big bang nucleosynthesis is not jeopardized by the decay of the gravitino gives a bound on the reheating temperature $T_{\rm R}\lesssim 10^5\sim 10^9$ GeV \cite{Kawasaki:2008qe}.
Extremely light, $m_{3/2}\lesssim$ eV, or extremely heavy, $m_{3/2}\gtrsim 10^7$ GeV \cite{Okada:2015vka}, gravitinos are unconstrained.
Although realizing such mass spectra in a realistic supersymmetry breaking mechanism is challenging, there exist possible scenarios, e.g. gravitino dark matter at $m_{3/2} \gtrsim$ EeV discussed in \cite{Dudas:2017rpa,Dudas:2017kfz,Garcia:2018wtq}.

%%%%%%%%%%%%%%%%%%%%%%%%%%%%%%%%%%%%%%%%%%%%%%%%
%\section{\label{sec:Constraints}
{\em Constraints from the reheating temperature.}---
Regardless of the details of the particle physics model that is embedded in supergravity, the constraints from the gravitino problem are always present.
The constraints give an upper bound on the reheating temperature.
It is thus important to elucidate the relation between the reheating temperature and the prediction for the cosmological parameters, whenever the viability of an inflationary model is discussed within supergravity.

Assuming the standard thermal history of the Universe, inflation (accelerated cosmic expansion) ends\footnote{
We use the condition that one of the slow roll parameters 
$\epsilon_V=(M_{\rm P}^2/2)(V_{{\rm E},\widehat\varphi}/V_{\rm E})^2$ or
$\eta_V=M_{\rm P}^2V_{{\rm E},\widehat\varphi\widehat\varphi}/V_{\rm E}$ reaches unity, namely, $\rm max(\epsilon_V,\eta_V)=1$ for the end of inflation.
This is in good agreement with the actual termination of accelerated cosmic expansion for the models studied here.}
at time $t_{\rm end}$,
followed by a period of (p)reheating characterized by the equation of state parameter $w$.
The Universe then thermalizes at time $t_{\rm th}$ and becomes radiation dominant\footnote{
Strictly speaking, the completion of thermalization and the start of radiation dominance (the end of reheating) are different, as emphasized e.g. in \cite{Harigaya:2013vwa}.
However, the distinction has little significance in our analysis due to the logarithmic dependance in the equation \eqref{eqn:Nk}.
We thus assume in our analysis that the Universe becomes radiation dominant immediately after thermalization.}
until matter-radiation equality is reached at time $t_{\rm eq}$. 
After that the Universe stays matter dominated, until today $t_0$.
The $e$-folding number $N_k$ between the horizon exit of the comoving wave number $k$ and the end of inflation is then expressed as \cite{Liddle:2003as,Martin:2010kz}
\begin{align}\label{eqn:Nk}
  N_k \equiv & \ln\frac{a_{\rm end}}{a_k}=66.5-\ln h-\ln\frac{k}{a_0H_0}
+\frac{1-3w}{12(1+w)}\ln\frac{\rho_{\rm th}}{\rho_{\rm end}}\crcr
&+\frac 14\ln\frac{V_k}{\rho_{\rm end}}
+\frac 14\ln\frac{V_k}{M_{\rm P}^4}
+\frac{1}{12}\left(\ln g_*^{\rm eq}-\ln g_*^{\rm th}\right),
%+\ln\frac{T_{\rm CMB}}{2.725\text{K}}
\end{align}
where $ H_0 = 100\,h\, {\rm km}\,{\rm s}^{-1}\,{\rm Mpc}^{-1}$ with $h=0.674$ \cite{Planck:2018vyg} is the Hubble parameter today,
$V_k$ is the potential \eqref{eqn:VE} evaluated at the time of the horizon exit of the wave number $k$, and
$a$, $\rho$, $g_*$ are the scale factor, the energy density and the number of relativistic degrees of freedom evaluated at the time specified by the super/subscripts ($k$ for the horizon exit, end for the end of inflation, th for the completion of thermalization (end of reheating), eq for the matter-radiation equality and 0 for the present time).

The equation of state parameter $w$ in \eqref{eqn:Nk} is understood to be the averaged value over the time $t_{\rm end}<t<t_{\rm th}$.
In the supergravity inflation scenario we consider, the inflaton has mass $M$ which is much smaller than the inflationary scale and is thus negligible during inflation.
Including this mass, the potential \eqref{eqn:VE} after inflation becomes
\begin{align}
  V_{\rm E}(\varphi)
  %=\frac{M_{\rm P}^4}{(M_{\rm P}^2+\xi\varphi^2)^2}
  %\left(\frac{y^2}{16}\varphi^4+\half M^2\varphi^2\right).
  \simeq
  \frac{y^2}{16}\varphi^4+\half M^2\varphi^2.
\end{align}
At the beginning of (p)reheating the quartic term dominates and the cosmic expansion is radiation-like, $w\simeq w_{\rm r}=1/3$.
As the amplitude of the inflaton oscillations is diminished, the quartic and the quadratic terms become comparable at time $t_\star$, when $\varphi=\varphi_\star\simeq\sqrt 8 M/y$.
Let us denote the energy density at this moment as $\rho_\star(<\rho_{\rm end})$. 
After $t_\star$, the quadratic term of the potential dominates and the cosmic expansion becomes matter-like, $w\simeq w_{\rm m}=0$.
Thus the (p)reheating of this model proceeds stepwise, first with radiation-like equation of state, and then with matter-like equation of state.
Accordingly, the fourth term of \eqref{eqn:Nk} may be written more concretely as
\begin{align}\label{eqn:EoSterm}
  &\frac{1-3w}{12(1+w)}\ln\frac{\rho_{\rm th}}{\rho_{\rm end}}\crcr
  &=\frac{1-3w_{\rm r}}{12(1+w_{\rm r})}\ln\frac{\rho_{\star}}{\rho_{\rm end}}
  +\frac{1-3w_{\rm m}}{12(1+w_{\rm m})}\ln\frac{\rho_{\rm th}}{\rho_{\star}}.
\end{align}
Now using $w_{\rm r}=1/3$, $w_{\rm m}=0$ and 
introducing dimensionless parameter $\delta$ ($0\leq\delta\leq 1$) to denote $\rho_\star = \delta^4\rho_{\rm end}$, \eqref{eqn:EoSterm} becomes
\begin{align}
  \frac{1}{12}\ln\frac{\rho_{\rm th}}{\rho_\star}
  =\frac{1}{12}\ln\left[\frac{\pi^2 g_*^{\rm th}}{30\rho_{\rm end}}\left(\frac{T_{\rm R}}{\delta}\right)^4\right].
\end{align}
Here, $T_{\rm R}$ is the reheating temperature and we have used 
$\rho_{\rm th}=\pi^2 g_*^{\rm th} T_{\rm R}^4/30$.
The reheating temperature always appears in the combination $T_{\rm R}/\delta$.
The energy density at the end of inflation may be evaluated as $\rho_{\rm end} \simeq 2\,V_{\rm end}$.
The parameter $\delta$ depends on the phenomenological model embedded in supergravity; 
for example, in the messenger inflation model \cite{Kawai:2021tzc} we find $\delta \sim 10^{-5}$ for the messenger mass $M = 10^8$ GeV and Yukawa coupling $y=5.735\times 10^{-6}$.

We solved the equations of motion for the supergravity inflation model to find the primordial tilt $n_s$ and the tensor-to-scalar ratio $r$, for given values of the reheating temperature $T_{\rm R}$.
The results are shown in Fig.~\ref{fig:nsrTR} as red curves, together with the 1- and 2-$\sigma$ contours from the Planck +BICEP/Keck 2018 data \cite{BICEP:2021xfz}.
The curves are found to be nearly straight lines with fitting formula
\begin{align}
  r=& 0.01-17.4\times (n_s-a_0-a_1 x-a_2 x^2 -a_3 x^3),\crcr
  & a_0 = 0.95935,\quad
  a_1 = 6.2000\times 10^{-4},\crcr
  & a_2 = -8.7565\times 10^{-6},\quad
  a_3 = 7.3869\times 10^{-8},
\end{align}
where $x\equiv\log_{10}(\delta^{-1}T_{\rm R}/\text{GeV})$.
We used %$g_*^{\rm eq}=3.36$ 
$g_*^{\rm eq}=3.91$ and $g_*^{\rm th}=106.75$ of the Standard Model.
%and chosen the pivot scale at $k=0.002\,{\rm Mpc}^{-1}$.
Generically, small inflaton mass is present over and above the quartic potential and the prediction depends on $\delta$ which parametrizes the transition between the radiation-like expansion and the matter-like expansion during (p)reheating.
%The pivot scale is chosen at 0.002 ${\rm Mpc}^{-1}$.
The highest reheating temperature admissible in supergravity inflation is $\sim 10^9$ GeV, and the lower bound of the reheating temperature compatible with big bang nucleosynthesis is a few MeV.
Since $\delta\leq 1$, the lower bound on the reheating temperature constrains the model to lie to the right of the leftmost red line of Fig.~\ref{fig:nsrTR}.
When $\delta = 1$, more than two thirds of the 1- and 2-$\sigma$ parameter regions on the $n_s$-$r$ plane are seen to be excluded by the gravitino constraints.
The 1-$\sigma$ bounds on the CMB observables give $\delta^{-1}T_{\rm R}<10^{31}$ GeV, and combining this with $T_{\rm R}\lesssim 10^9$ GeV we have a model-independent lower bound on the parameter $\delta>10^{-22}$.
The steep slope of the red lines indicates strong correlation between the rescaled reheating temperature $\delta^{-1}T_{\rm R}$ and the primordial tilt $n_s$.
%susceptible to the bounds.
Thus, measurements of $n_s$ are important to test this class of inflationary scenarios.
In future, precision measurements of $n_s$ combined with the constraints from the gravitino problem may well rule out this otherwise promising model of inflationary cosmology.
% ('_')b

%Figure \ref{fig:nsrTR} shows the 

%%%%%%%%%%%%%%%%%%%%%%%%%%%%%%%%%%%%%%%%%%%%%%%%
%\section{\label{sec:Final}
%Final remarks}

{\em Note Added.}---
After completing this work, we noticed a preprint 
\cite{Cheong:2021kyc} with partially overlapping results appeared on the arXiv.

%%%%%%%%%%%%%%%%%%%%%%%%%%%%%%%%%%%%%%%%%%%%%%%%
%\begin{acknowledgments}
{\em Acknowledgments.}---
This work was supported in part by the National Research Foundation of Korea Grant-in-Aid for Scientific Research Grant No.
NRF-2018R1D1A1B07051127 (S.K) and by the United States Department of Energy Grant No. DE-SC0012447 (N.O.).
%\end{acknowledgments}

%%%%%%%%%%%%%%%%%%%%%%%%%%%%%%%%%%%%%%%%%%%%%%%%

%%%%%%%%%%%%%%%%%%%%%%%%%%%%%%%%%%%%%%%%%%%%%%%%
%\appendix
%\section{Appendix A}
%%%%%%%%%%%%%%%%%%%%%%%%%%%%%%%%%%%%%%%%%%%%%%%%

%%%%%%%%%%%%%%%%%%%%%%%%%%%%%%%%%%%%%%%%%%%%%%%% 
%\bibstyle{h-physrev.bst} 
%\providecommand{\href}[2]{#2} 
%\bibliography{C:/Users/Shinsuke/Dropbox/Archive/BibTeX/AdSCFT.bib,C:/Users/Shinsuke/Dropbox/Archive/BibTeX/pheno.bib} 
%\bibliography{/Users/kawai/Dropbox/Archive/BibTeX/AdSCFT.bib,/Users/kawai/Dropbox/Archive/BibTeX/pheno.bib} 
\input{SugraInfBK18_4.bbl}

%%%%%%%%%%%%%%%%%%%%%%%%%%%%%%%%%%%%%%%%%%%%%%%%
\end{document}